# Quantum dots emission enhancement via coupling with an epsilon-near-zero sublayer

S. Stengel,[1] A. B. Solanki,[2,3] H. Ather,[4,2] P. G. Chen,[2] J. I. Choi,[2] B. M. Triplett,[2] M. Ozlu,[2] K. R. Choi,[2,5] A. Senichev,[2] W. Jaffray,[1] A. S. Lagutchev,[2] L. Caspani,[6] M. Clerici,[6] L. Razzari,[7] R. Morandotti,[7] M. Ferrera,[1] A. Boltasseva,[2] and V. M. Shalaev[2,4,3,8,*]

[1] Institute for Photonics and Quantum Science, Heriot-Watt University, Edinburgh EH14 4AS, UK
[2] Birck Nanotechnology Center, Purdue University, West Lafayette, Indiana 47907, USA
[3] Elmore Family School of Engineering and Computer Engineering, Purdue University, West Lafayette, Indiana 47907, USA
[4] Department of Physics and Astronomy, Purdue University, West Lafayette, Indiana 47907, USA
[5] Research Institute for Nanoscale Science & Technology, Chungbuk National University, Cheongju, Chungbuk, 28644, Republic of Korea
[6] Como Lake Institute of Photonics, Department of Science and High Technology, University of Insubria, Via Valleggio 11, 22100, Como, Italy
[7] Institut National de la Recherche Scientifique Centre Énergie Matériaux Télécommunications (INRS-EMT), 1650 Boulevard Lionel-Boulet, Varennes, Quebec J3X 1P7, Canada
[8] Purdue Quantum Science and Engineering Institute (PQSEI), Purdue University, West Lafayette, Indiana 47907, USA

(Dated: 11 September 2025)

Quantum emitters operating at telecom wavelengths are essential for the advancement of quantum technologies, particularly in the development of integrated on-chip devices for quantum computing, communication, and sensing. Coupling resonant structures to a near-zero-index (NZI) environment has been shown to enhance their optical performance by both increasing spontaneous emission rates and improving emission directionality. In this work, we comparatively study emission characteristics of colloidal PbS/CdS (core/shell) quantum dots at telecom wavelengths on different substrates, where two different sets of quantum dots emitting within and outside the epsilon-near-zero region are deposited on both glass and indium tin oxide (ITO) substrates. Our results demonstrate that coupling quantum dots to the epsilon-near-zero spectral region results in a reduction of photoluminescence lifetime of 54 times, a 7.5-fold increase in saturation intensity, and a relative emission cone narrowing from 17.6° to 10.3°. These results underline the strong dependence of quantum dot emission properties on the spectral overlap with the epsilon-near-zero condition, highlighting the potential of transparent conducting oxides (TCOs), such as ITO, for integration into next-generation quantum photonic devices. Due to their CMOS compatibility, fabrication tunability, and high thermal and optical damage thresholds, TCO NZI materials offer a robust platform for scalable and high-performance quantum optical systems operating within the telecom bandwidth.

## I. INTRODUCTION

In recent years, the field of quantum optics has witnessed a surge in applications that were once considered technologically unfeasible. Among these, solid-state quantum emitters have emerged as versatile platforms for single-photon generation, enabling advancements in quantum sensing, secure communication, and integrated quantum photonic circuits[1]. In particular, emitters operating in the telecom wavelength band are key to merging quantum functionalities with existing fiber-optic infrastructures[2]. This has catalyzed efforts towards integrating quantum emitters with an on-chip photonic environment to control emission dynamics, device complexity, and coherence[3,4]. PbS/CdS core–shell quantum dots (QDs) are a compelling choice for these applications, thanks to their commercial availability, size-tunable emission in the near-infrared (NIR), and compatibility with non-cryogenic operation[5]. Their spectral range spans from 850 nm to 1600 nm depending on size and ligand, with improved photostability and passivation due to the CdS shell[6]. These QDs exhibit native lifetimes on the order of 1–3 μs and maintain relatively good quantum efficiency over a wide temperature range[5,7]. The surface chemistry of QDs can also be tuned to control radiative vs. non-radiative decay pathways[8], and their utility has been demonstrated in diverse areas from solar energy harvesting[9,10] to bioimaging[11,12]. However, their integration into a telecom photonic environment that could reshape and enhance their emission properties remains an open challenge.

A promising route to enhance and control spontaneous emission lies in the possibility of coupling quantum emitters to the so-called epsilon-near-zero (ENZ) materials, for which the real part of the dielectric permittivity vanishes near the plasma frequency, while the imaginary part is relatively small (such materials are also referred to as near-zero-index, NZI, materials). Transparent conducting oxides (TCOs) such as indium tin oxide (ITO) exhibit ENZ behavior in the NIR spectral region, are CMOS compatible, and allow spectral tuning via doping and deposition conditions[13,14]. Near their ENZ point, these materials exhibit a number of unusual optical effects like out-of-plane electric field suppression, and strong nonlinear responses[15–18]. Furthermore, the radiative properties of emitters in and around ENZ materials are dependent on the materials' dimensionality, thus providing a higher degree of freedom in engineering dispersion and associated emission properties[3,19–21]. Theoretical works also predict that coupling emitters to an ENZ environment can lead to a drastic reshaping of their radiation properties, including radiative suppression, modification of the emission directionality, and even enabling super-radiant behavior[19,22–25]. In addition to this, the optical analogue of the Meissner effect, where ENZ substrates prevent field penetration, provides an additional degree of freedom to engineer emission properties in a low index systems[26,27]. Also, by placing quantum emitters in a time-modulated ENZ environment, further control of quantum states could be achieved for the design of fully optical quantum networks[28,29]. Studies also confirm that nanoantennas placed on ENZ substrates exhibit enhanced out-of-plane radiation and narrower angular emission when the spectral overlap with the ENZ condition is met[27].

*)corresponding author: shalaev@purdue.edu

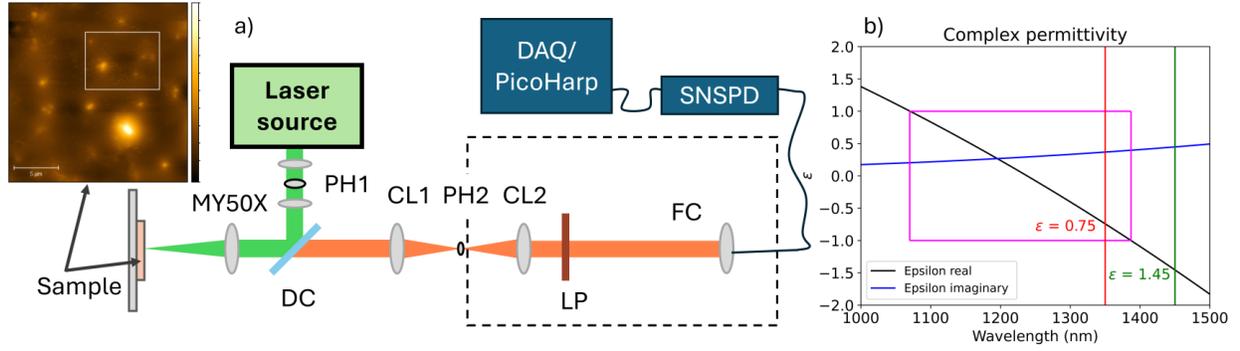

FIG. 1. Experimental setup. The figure illustrates the simplified excitation/detection scheme for both saturation and lifetime measurements. a) At first, the laser beam centered around 530 nm passes through a confocal pinhole (PH1), then reflects off a dichroic mirror (DC), passes through a microscope objective (MY50X), which is mounted on a 3D piezo stage, and is then focused onto the sample (Fig. 1(a) inset shows sample AFM image). The emitted signal passes the DC and is directed through PH2 (200 μm) and lenses CL1 and CL2 for spatial filtering. All components after PH2 are housed inside a blacked-out box, followed by a long-pass filter (LP) with a cut-on wavelength at 1 μm, omitting any stray light and allowing only the emitters' signal to be detected. Further, the beam is then fiber-coupled at (FC), collecting the signal and sending it to the superconducting nanowire single-photon detector (SNSPD) with high efficiency in the near-infrared. The SNSPD's electronic readout is passed to either a data acquisition device (DAQ) for counts evaluation or to the PicoHarp module for lifetime measurements. b) Dispersion curves (epsilon real for the black curve and epsilon imaginary for the blue curve) of ITO, with its ENZ bandwidth (1070 nm to 1380 nm ($|\varepsilon'| < 1$), pink box), and the emission of both QD batches at 1350 nm (red vertical line) and 1450 nm (green vertical line).

To date, a direct and broad experimental exploration of room-temperature emission properties of ENZ-coupled QDs within the telecom band remains sparse. In this work, we present a comparative study of colloidal PbS/CdS QDs emission properties when deposited on standard microscope glass slides and on an ITO thin film. Our analysis also considers cases where the QD emission wavelength falls within and outside the ENZ band of the ITO substrate. Within these experimental settings, we observed a 54-fold reduction in lifetime, a 7.5-fold increase in photoluminescence (PL) intensity, and a narrowing of the emission cone from 17.6° to 10.3°. These results are consistent with prior observations of radiation engineering in the ENZ environment and pave the way for scalable, CMOS-compatible platforms based on ENZ-to-QD coupling.[8,19,22,25,27].

## II. EXPERIMENTS AND DISCUSSIONS

### A. Experimental setup

To investigate the emission characteristics of PbS/CdS QDs deposited on glass and ENZ ITO thin film, we employed a custom-built time-correlated single-photon counting (TC-SPC) system optimized for detection in the NIR telecom range, while simultaneously allowing for excitation in both the visible and NIR ranges. The core of the system is a confocal microscope combined with the TCSPC setup as depicted in a simplified schematic in Fig. 1(a). This setup enables the measurement of lifetime, emission directionality, and saturation behavior. Initially, the laser beam emitting around 530 nm passes through a spatial filter (PH1) to achieve the smallest possible excitation spot size. Then the beam reflects off a dichroic mirror (DC), with 98% reflection efficiency around the green standardized excitation wavelength, while allowing 70% transmission for the NIR emission. The excitation beam is focused onto the sample using a 0.42NA 50× Mitutoyo Plan Apo NIR infinity-corrected microscope objective (MY50X), resulting in a focal spot smaller than 5 μm (full width half maximum). This was experimentally confirmed by recording the PL images from a series of calibrated grating patterns. Confocal scanning of the sample was performed using an objective mounted on an X-Y-Z piezo stage, which allowed for sub-nanometer positioning precision while the sample remained stationary. The emitted PL was collected through the objective and imaged onto a confocal pinhole (PH2, CL1, and CL2). All optics following PH2 are enclosed in a blacked-out box to eliminate both ambient light and residual pump,

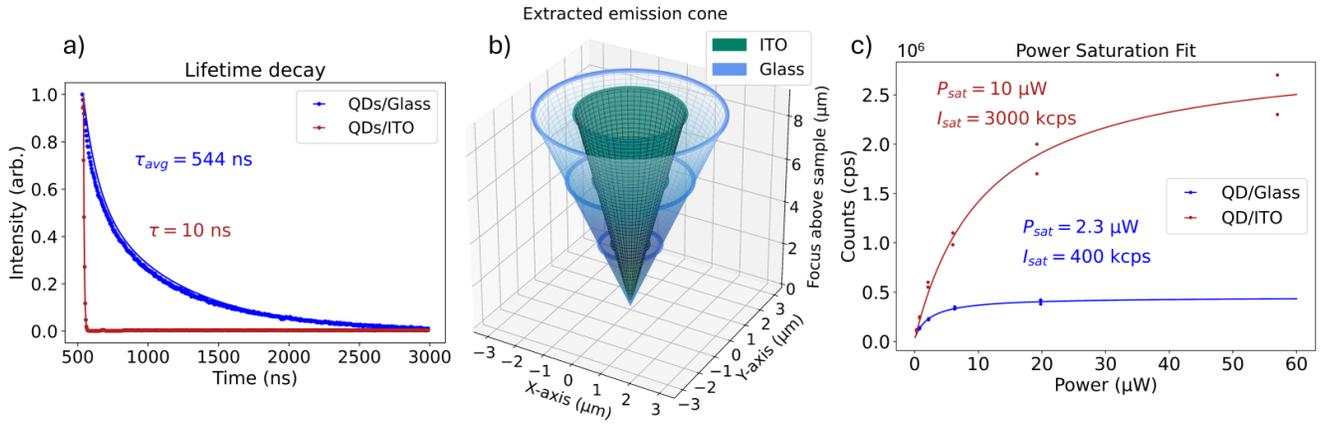

FIG. 2. Comparative analysis of QDs emitting at 1350 nm (within the ENZ bandwidth of ITO) on glass and ITO. In (a), a significant reduction of lifetime is observed, from 544 ns on glass (blue line and markers) to 10 ns when coupled to an ENZ environment (red line and markers). Panel (b) illustrates the change in directionality, with a relative emission cone narrowing from 17.6° to 10.3° for QDs on glass (blue) and on ITO (green) substrate, respectively. In (c), the saturation behavior of QDs reveals an increase in saturation power from $P_{sat}$ =2.3 µW to 10 µW, alongside a 7.5-fold enhancement in saturation intensity $I_{sat}$, from 400 kcps to 3000 kcps, when comparing QDs on ITO (red line and markers) with those on glass (blue line and markers)

and to ensure signal collection from the focal plane only[30]. Additionally, inside the enclosed box, a 1 µm long-pass filter was installed to ensure that only the PL from the sample was fiber-coupled and sent to the superconducting nanowire single-photon detector (SNSPD) (see Fig. 1(a)). The SNSPD output was connected to a custom-built microelectronic interface that conditions and routes the electrical signals to either a data acquisition device (DAQ, count rate evaluation) or a PicoHarp (lifetime measurement) TCSPC module. The setup was controlled, and data was acquired via a LabVIEW code. PL maps were recorded by scanning the piezo stage with the attached microscope objective, while synchronously collecting the count rates. Subsequently, the PL maps were used to identify the most suitable regions containing spatially isolated QD ensembles. Once an emission region was identified, the microscope's objective was focused onto it to perform the lifetime measurement. For the lifetime measurements, pulses centered in the green with a full-width half-maximum duration of less than 20 ns were used to excite the QDs.

### B. Sample preparation

The QDs employed in this study, sourced from CD Bioparticles, feature a commonly used PbS/CdS core/shell structure. We utilized two distinct batches of QDs, where one has a PL peak emission at 1350 nm, aligned with the ENZ bandwidth of the ITO sublayer, and the other one peaks at 1450 nm, which lies outside this ENZ region. To ensure optimal distribution across the substrate, the QDs were diluted in acetone at a 10:1 ratio of acetone to QDs before drop-casting. Two types of substrates were prepared to comparatively study the emission properties of the QDs. The reference substrate was a standard microscope glass slide, and the other substrate was made of TCO, specifically from a ∼ 240 nm thick ITO thin film (dispersion curves shown in Fig. 1(b)). Both substrate types underwent chemical cleaning, including sequential immersion in isopropanol, methanol, and acetone for 30 seconds each, including nitrogen blow-drying. Subsequently, the substrates were subjected to 90 minutes of UV-C light exposure to enhance surface wettability and promote a uniform QD dispersion. To prevent aggregation and achieve a highly uniform coating, the diluted QD solution was sonicated in an ultrasonic bath for ∼ 45 minutes just prior to deposition. Following this, the solution was drop-cast onto the cleaned substrates and placed in vacuum storage boxes to accelerate solvent evaporation, ensuring an even spread of QDs. An atomic-force-microscopy (AFM) image is depicted as an inset in Fig. 1(a), showing the distribution of QD ensemble over a select 55 µm² area. Importantly, the ENZ bandwidth of the ITO substrates spans from 1070 nm to 1380 nm ($|\varepsilon'| < 1$), overlapping with the emission band of the first QD batch, portrayed in Fig. 1(b). Conversely, the second batch, which emits at 1450 nm, is strategically positioned outside the ENZ bandwidth of the ITO thin film, allowing for a comparative study, see Fig. 1(b).

### C. Results and discussion

#### 1. Emission within the ENZ region

We first examine the emission characteristics of PbS/CdS QDs whose emission spectrum falls within the ENZ region of the ITO substrate. A pronounced reduction in PL lifetime from 544 ns to 10 ns is observed for QDs drop-cast on the ITO substrate compared to those on glass, as reported in Fig. 2(a). In this regard, we should mention that our minimum time resolution is set by the pulse duration, setting only an upper limit to the fastest measurable decay time. The lifetime on glass is calculated using a bi-exponential fit and is averaged using an intensity-weighted formulism, according to the following equations (1) and (2)[31]:

$$I_{PL}(t) = Ae^{-\frac{t}{\tau_1}} + Be^{-\frac{t}{\tau_2}}, \quad (1)$$

$$\tau_{avg} = \frac{A\tau_1^2 + B\tau_2^2}{A\tau_1 + B\tau_2}, \quad (2)$$

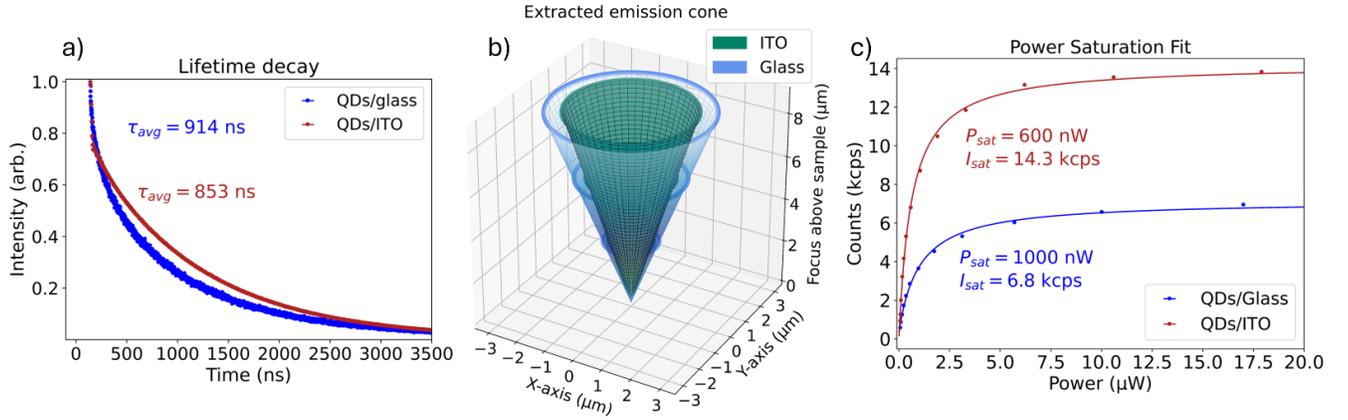

FIG. 3. Comparative analysis of QDs emitting at 1450 nm (outside the ENZ bandwidth of ITO) on glass and ITO. In (a), a minimal change of less than 10% is observed for the emitter lifetime between the cases of QDs on glass (blue line and markers) and on ITO (red line and markers), yielding 914 ns and 853 ns, respectively. Panel (b) shows the emission cones for QDs on both substrates, with a slightly narrower emission cone of 12.8° for the ITO case (green) and a wider cone of 16.0° for the QDs on glass (blue). Panel (c) depicts the saturation behavior outside the ENZ region and reveals a lower saturation power $P_{sat} = 600$ nW vs. 1000 nW, and a higher saturation intensity $I_{sat} = 14.3$ kcps vs. 6.8 kcps, when comparing QDs on ITO (red line and markers) and glass (blue line and markers), respectively.

where $I_{PL}(t)$ is the fitted PL decay, $\tau_{avg}$ is the intensity-weighted-average lifetime, $A$ and $B$ are the weights, and $\tau_1$ and $\tau_2$ are the corresponding bi-exponential decays. The lifetime of QDs on ITO is calculated using a single exponential decay mode, due to the rapid decay and the limited number of data points. It is worth mentioning that the QDs on glass exhibited a lifetime decay similar to other reported values for PbS/CdS core/shell QDs[5,6,8], and that coupling to the ENZ thin-film reduces the lifetime by 54 times to 10 ns. At this point, we should highlight that despite the advanced control behind QD fabrication processes, their use in integrated photonics has been primarily limited by a long decay time, impeding key applications such as high-repetition-rate photons on demand. Our results directly address this critical drawback.

From the point of view of the photon directionality, emission profiles were reconstructed by collecting the PL intensity at different focal planes at 0, 3, 6, and 9 μm on and above the sample surface. This was done while maintaining a constant excitation power density over the scanning distance, with a variation of about ±3% on the peak intensity. The collected data were then fit with a standard Gaussian beam divergence model to extract the angular spread. This analysis resulted in the distinct radiation signatures, as illustrated in Fig. 2(b), where the emission cone of QDs decreases from 17.6° on glass to 10.3° on ITO. Finally, to characterize the saturation behavior, the count rate of the QD ensemble was monitored as the pump power was gradually increased. The saturation curves were fitted using the standard three-level emitter saturation model for optical emitters[32,33],

$$I(P) = I_{sat} \frac{P}{P + P_{sat}}. \quad (3)$$

Here, $I(P)$ is the emission intensity as a function of the excitation power $P$, the maximum achievable saturation intensity is $I_{sat}$ [cps], and $P_{sat}$ [μW] is the power at which the system saturates. These plots are shown in Fig. 2(c), where, due to the ENZ coupling, the QDs on ITO perform significantly better than the ones on glass, in terms of both the saturation count rate and the saturation power. Specifically, these parameters were recorded to be $I_{sat} = 3000$ kcps and 400 kcps, and $P_{sat} = 10$ μW and 2.3 μW, for QDs on ITO and glass, respectively. This combined analysis confirms the strong interaction between quantum emitters and an epsilon-near-zero environment, as reflected in their modified emission lifetime, directionality, and saturation response. Our study also complements previous work on visible QDs coupled to an ENZ environment[3].

2. Emission outside the ENZ region

In order to disentangle the emission enhancement due to the ENZ coupling from any other effect due to the presence of the ITO substrate, we performed an equivalent comparative study as the one previously reported, this time with QDs emitting at 1450 nm, outside the ENZ bandwidth. As before, measurements were focused on PL lifetime, emission directionality, and saturation behavior. For these cases, lifetimes of $\tau_{avg} = 914$ ns and $\tau_{avg} = 853$ ns have been recorded for QDs on glass and ITO, respectively. Thus, it can be seen that lifetime reduction due to the ITO substrate is still visible, but is considerably less pronounced when compared to the previous ENZ-coupled case.

The emission directionality measurements (Fig. 3(b)) show that the emission cone of QDs on ITO (green) is 12.8°, which is only slightly narrower than the 16° emission cone of QDs on glass (blue), indicating a relatively small directional enhancement.

Finally, we analyzed the power-dependent saturation behavior for QDs on both substrates, and the count rate as a function of input power (Fig. 3(c)). Here, the QDs on ITO exhibit a

Quantum dots emission enhancement via coupling with an epsilon-near-zero sublayer  5Quantum dots emission enhancement via coupling with an epsilon-near-zero sublayer 5 slightly higher saturation count rate, $I_{sat}$, of 14.3 kcps as compared to 6.8 kcps on glass. Further, the saturation power for the QDs on ITO and glass are $P_{sat} = 600$ nW and 1000 nW, respectively. This analysis shows that the enhancement in the emission properties of QDs coupled to ITO is primarily due to their emission overlap with the ENZ region. The minor enhancement effect on all investigated parameters outside of the ENZ bandwidth can be ascribed to plasmonic enhancement (Purcell effect). Specifically, at 1450 nm, ITO exhibits more metallic properties, thus increasing the reflectivity and contributing to the higher saturation intensity. Another factor at play is the refractive index contrast, which supports a narrower cone and therefore a more directed emission, thereby boosting the count rate at equal input power for an ITO substrate at 1450 nm compared to glass.

We should note that the observed lifetime reduction does not correspond one-to-one to an enhanced count rate for all investigated cases. This can be explained by a redistribution of radiative and non-radiative modes, and the uncertainty in the number of excited emitters within the studied ensemble. To deepen these aspects, future studies employing sub-ps pulses of different time durations may allow for the identification of multiple recombination processes in action. Alongside these, it would be very interesting to refine our study at the single- and few-photon level. However, these investigations are outside the scope of our phenomenological study.

## III. CONCLUSION

This study presents clear experimental evidence of the profound influence that coupling to an ENZ environment can have on quantum emitter properties. When PbS/CdS (core/shell) QD emission falls within the ENZ bandwidth of the ITO-underlayer, a nearly two-order-of-magnitude reduction in the photoluminescence lifetime from 544 ns down to 10 ns is demonstrated when compared to QDs on glass. Additionally, a 7.5-fold increase in the saturation intensity $I_{sat}$ and a 4-fold increase in the saturation power $P_{sat}$ were observed, suggesting a significantly more efficient excitation-emission cycle. Furthermore, the angular emission profile narrowed significantly, from 17.6° to 10.3°, indicating a considerably improved directionality.

These enhancements are caused by the presence of the ENZ condition, as demonstrated by the comparative study performed with QD emission outside the ITO's ENZ bandwidth. Our work aims to stimulate discussions on the potential impact of low-index materials on integrated quantum technologies, motivating further studies towards single-photon emission and possible superradiant effects. The observed results clearly indicate that the ENZ environment provides a highly effective platform for controlling light-matter interactions, enabling significant improvements in emission efficiency and directionality, all within a CMOS-compatible, telecom-band platform. Such capabilities are crucial for advancing emerging on-chip quantum optics, low-threshold quantum light sources, and tunable emitter arrays in scalable quantum photonic networks.